\documentclass[aps,prl,twocolumn,groupedaddress,showpacs]{revtex4}
\usepackage{epsf,color}

\usepackage[dvips]{epsfig}

\scrollmode
\begin{document}


\title{Exciton photon strong-coupling regime for a single quantum dot in a microcavity.}



\author{E. Peter, P. Senellart, D. Martrou, A. Lema\^{\i}tre, J. Hours, J.M. G\'erard$^{\dag}$ and J. Bloch}
\affiliation{CNRS-Laboratoire de Photonique et Nanostructures,
Route de Nozay, 91460 Marcoussis, France
\\$^{\dag}$CEA/DRFMC/SP2M, Nanophysics and Semiconductor
Laboratory, 17 rue des Martyrs, 38054 Grenoble Cedex, France }

\email[]{emmanuelle.peter@lpn.cnrs.fr}


\date{\today}

\begin{abstract}
We report on the observation of the strong coupling regime between a
single GaAs quantum dot and a microdisk optical mode.
Photoluminescence is performed at various temperatures to tune the
quantum dot exciton with respect to the optical mode. At resonance,
we observe an anticrossing, signature of the strong coupling regime
with a well resolved doublet. The Vacuum Rabi splitting amounts to
$400\, \mu eV$  and is twice as large as the individual linewidths.
\end{abstract}


\maketitle


Cavity Quantum Electrodynamics (CQED) has motivated a lot of
fascinating experiments in Atomic Physics these last twenty years
\cite{Haroche}. The spontaneous emission of light from atoms
inserted in a microcavity can be controlled \cite {Goy} and even
become reversible in the so-called strong coupling regime (SCR)
using high finesse cavities \cite {Thompson}. More recently,
analogous CQED approach has been applied to the discrete states of
semiconductor quantum dots (QD). The control of spontaneous emission
with QDs inserted in microcavities has been observed \cite  {JMG,
Gayralapl78,Grahamapl74,Solomon,Kiraz,Moreau1} and applied to the
realization of efficient single photon sources \cite {Moreau1,
Santori}.  As in atomic physics \cite {Hood}, the SCR with an
atomic-like QD state is of fundamental interest for new CQED
experiments \cite{McKeever} but also for new solid state devices
such as single QD lasers \cite {MichlerJMG} or quantum logical gate.
Indeed the discrete QD states could constitute the elementary
building block of the solid-state quantum computer (referred to as
qubits)  \cite {Michler,Molotkov,Troiani} and the electromagnetic
field of the cavity mode would mediate the interaction between
qubits \cite {Imamoglu,Pellizari}.

 When inserting a single QD inside an optical
microcavity, two regimes can be reached, depending on the coupling
strength between the QD exciton and the optical cavity mode \cite
{Fabre}. In the weak coupling regime, the spontaneous emission rate
of the QD exciton is modified as compared to outside the cavity.
This phenomenon referred to as Purcell effect \cite {Purcell} has
been observed for a QD inside a cavity \cite {Gerard,Graham,Bayer}.
In the SCR, the exciton-photon coupling is stronger so that the
spontaneous emission becomes reversible. Photons emitted by the QD
inside the cavity mode are re-absorbed, re-emitted... giving rise to
the so-called one-photon Rabi oscillations. This coherent evolution
takes place as long as the Rabi oscillation is faster than the
decoherence due to both the exciton and the cavity mode. Until now,
the SCR has not
 been observed for single QD in optical microcavity, either because the cavity losses are too
 large or the oscillator strength of the investigated QD (most of the times, InAs self-assembled QDs)
 is too small \cite{Kiraz}.

In the present letter we report on the first experimental
observation of the strong coupling regime between a single QD and an
optical microcavity mode. Using temperature tuning, we observe the
spectral signature of mixed exciton photon states by
photoluminescence. The quantum dots we use are quantum dots formed
at the interface fluctuations of a thin GaAs quantum well. As
theoretically predicted \cite {Andreani} and checked experimentally
\cite {Stievater,Hourssubmit}, these monolayer fluctuation QDs have
an oscillator strength much larger than InAs self-assembled QDs. For
this reason, Andreani and co-workers have predicted that the SCR
could be achieved when these QDs are inserted in a state-of-art
micropillar. Here, we use another type of cavities, which present
optical modes with even higher quality factors: microdisks supported
by a small pedestal surrounded by air \cite {Gayral}. Within such
microdisks, whispering gallery modes can establish: they are
vertically confined by the large index contrast between
semiconductor and air; they are guided in the circumference of the
disk by total internal reflection \cite {Mccall}.

Our sample was grown by molecular beam epitaxy on a GaAs
substrate. Growth conditions were optimized to obtain large QDs,
which are expected to present large oscillator strength \cite {Andreani}.
After the buffer layer, a 10 minute annealing at 640
$^{\circ}$C was performed under arsenic flow. Then a 1.5 $ \mu m$
layer of $Al_{0.8}Ga_{0.2}As $ was grown at low temperature
(555$^{\circ}$C). The growth temperature was raised to
600$^{\circ}$C and a 71 nm short-period superlattice of 50 $\%$
mean Al composition was grown. Then, the active material was
deposited: a 50 $nm Al_{0.33}Ga_{0.67}As $ barrier followed by a
13 monolayer (around 3.7 $nm$) GaAs QW was deposited, at a growth
rate of 1.5 (resp. 1.0) \AA .s$^{-1}$ for the barrier (resp. QW).
The growth temperature was then lowered to 590$^{\circ}$C before
depositing the top barrier (similar to the first one) to decrease
element III interdiffusion. Finally the same 71 nm superlattice
was grown on top of the structure to make it symmetric. A 120
s growth interruption was performed at each QW interface to
smoothen the growth surface. We used a Scanning Tunneling
Microscope (STM) coupled to the growth chamber to study the
surface morphology at each interface. Figure 1c and 1d present STM
picture of both interfaces just after the growth interruption. At
the first interface (fig. 1c), the surface is slightly rough with
a few holes. On the opposite, after growth of 13 GaAs monolayers,
the surface has become atomically smooth with no islands. In this sample, a
 monolayer fluctuation in the quantum well thickness resulting in a
hole in the first interface $Al_{0.33}Ga_{0.67}As / GaAs$, or in
an island in the second interface, creates an attractive potential
(quantum dot) that can localize the centre of mass of the quasi
2D-exciton confined in the quantum well. Analysis of these STM
images shows that the lateral size of the quantum dot is between
100 and 1 000 nm and that its areal QD density is around 10 $\mu m ^{-2}$.

\begin{figure}[h]
\begin{center}
\epsfig{file=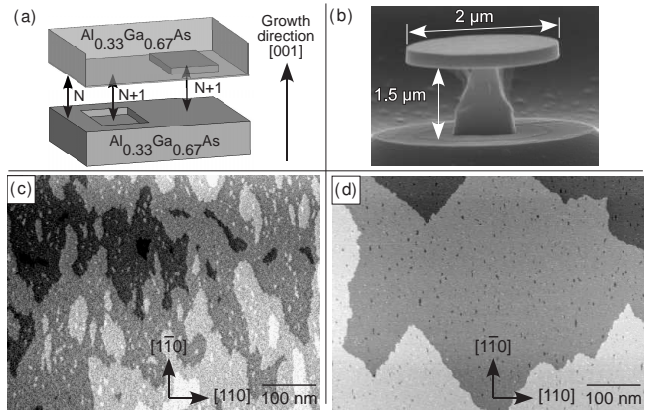, width= 8.5 cm}
\end{center}
\caption{\small { (a) Schematic of a monolayer fluctuation QD.
(b)Scanning electron microscopy sideview of a 2 $\mu m $ diameter
microdisk. (c) $800 nm*600 nm$  Scanning tunnelling microscopy (STM)
image of the surface after the deposition of the first 50 $nm
Al_{0.33}Ga_{0.67}As $ barrier and after the growth interruption;
the color changes correspond to one (AlGa)As monolayer changes in
height (d), STM image of the surface after growth of the 13
monolayer GaAs QW and after the growth interruption.}}
\end{figure}

  To realize the optical microcavity, we performed an electron beam lithography followed by the lift-off of a 10 nm Ti mask in order to pattern an array
   of 3.8 $ \mu m$ diameter circular mesas. Then the sample was chemically etched by a non selective quasi isotropic solution (dichromate-
   based solution). After a vertical etching of 1.5 $\mu m $, the diameter
of the disk is reduced by roughly 1.8 $ \mu m$. Then, to define the
microdisk pedestal, we perform a selective etching of the 1.5 $\mu
m$ Al rich layer using a diluted HF solution \cite {Gayral}. This
last step also removes the Ti mask. Figure 1b shows a side-view of a
typical 2 $\mu m$ diameter microdisk constituting the optical cavity
with the active layer inside.

Photoluminescence measurements are performed at a cryogenic
temperature using a cold-finger helium cryostat. The excitation beam is delivered by a continuous wave Ti:Sapphire
laser (of energy 1750 meV) focused with a microscope objective (numerical aperture 0.5) onto a 2 $ \mu m$ diameter
excitation spot. The emission, collected at normal incidence by the same objective, is dispersed by a double grating spectrometer and
detected with a $N_{2}$-cooled Si-CCD camera with a 80 $\mu eV$ spectral resolution.

\begin{figure}[h]
\begin{center}
\rotatebox{270}{\epsfig{file=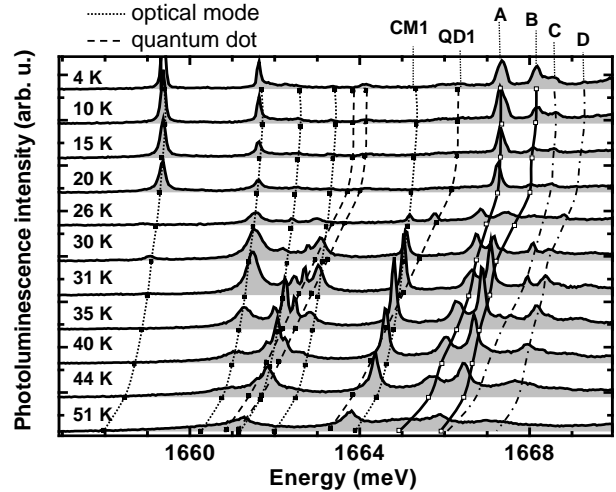, height = 8 cm}}
\end{center}
\caption{\small { Photoluminescence spectra (vertically shifted for
clarity) on a single microdisk at various temperaturesfrom 4K to 51
K.  The dotted (resp. dashed) lines are guides
 to the eye to follow the emission energy of the optical modes (resp. quantum dot excitons). Solid lines
 follow spectral shifts different both from a quantum dot
  and an optical mode spectral shift.}}
\end{figure}

Figure 2 shows the spectra obtained on a single microdisk for
various temperatures, between 4 K and 51 K.  The observed
luminescence peaks correspond either to the emission of quantum dots
or to the emission of a small background within an optical cavity
mode.  Varying temperature allows identifying QDs versus optical
modes. As previously reported \cite {Kiraz}, quantum dot emission
energy exhibits a stronger spectral shift with temperature than an
optical mode. The redshift of the QD when heating the sample is due
to changes in the GaAs and $Al_{0.33}Ga_{0.67}As $ band gaps,
whereas the redshift of the optical mode is due to the change of the
microdisk AlGaAs index. The dashed lines (resp. dotted lines) on
figure 2 follow the QD (resp. optical mode) emission energy. Because
of this difference in temperature dependence, we can tune a QD
exciton with respect to a cavity mode by changing the sample
temperature. On figure 2, we observe seven whispering gallery modes
within a 15 meV spectral window. This is consistent with the number
of modes expected for a 2 $\mu m$ diameter microdisk with 250 nm
thickness at this wavelength.

Let us now consider the two lines labelled QD1 and CM1 on figure 2.
At low temperature, QD1 is on the high-energy side of the cavity
mode CM1. When increasing the temperature, the QD red-shifts more
than the optical mode as is particularly visible at 26 K. Then a
strong increase of the emission within the cavity mode CM1 is
observed around 31 K, temperature where QD1 and CM1 are in
resonance. This QD is in the weak coupling regime with the cavity
mode. At resonance, because of the Purcell effect, the QD emits
photons preferentially within this mode. As a consequence, the
emission intensity at the cavity mode energy is enhanced \cite
{Gerard,Graham,Bayer}. This intensity evolution with temperature
would be completely different for a quantum dot not coupled to an
optical mode: when increasing temperature, the emission also
redshifts but simultaneously shows a progressive quenching of
intensity because thermal heating results in exciton escape out of
the quantum dot \cite{Peterphonon}. Notice also that around 1662
meV, we observe the resonance between two other QDs and two cavity
modes.

\begin{figure}[h]
\rotatebox{270}{\epsfig{file=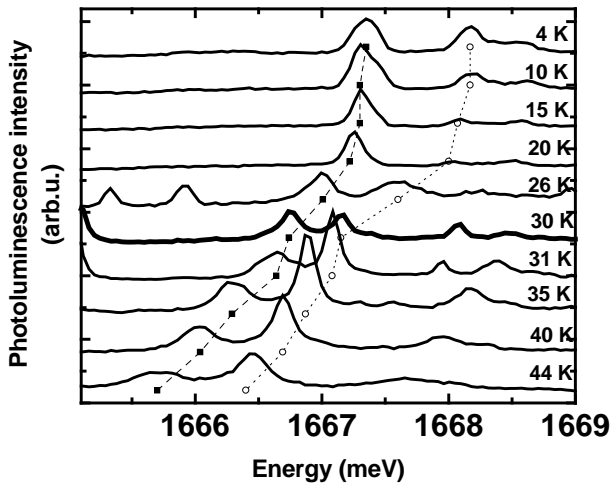, height = 8 cm}}
  \caption{ \small {Photoluminescence spectra (vertically shifted for clarity) for various
 temperatures between 4 K and 44 K. Dashed circles and squares: guide to the eye to follow the energy
  of the emission peaks for each temperature.}}
\end{figure}

Consider now the emission peaks A and B on figure 2: this spectral
shifts with temperature differ both from the typical exciton or
optical mode behavior. A zoom of this region is presented on Figure
3. When increasing temperature from 4 K up to 30 K, the upper line
shifts more than the lower line. This means that the upper line  can
be assigned to the quantum dot emission, and the lower line to a
cavity mode. The emission linewidth of the lower line amounts to 0.2
meV, corresponding to a quality factor of 8000 for the considered
whispering gallery mode. The emission linewidth of the upper line
also amounts to 0.2 meV. This homogeneous broadening is not
radiatively limited but due to dephasing mechanisms, such as phonon
interaction \cite {Peterphonon, Borri}. When increasing temperature,
we do not observe a crossing as with QD1 and CM1, but an
anticrossing, signature of the Strong Coupling Regime. Above 31 K,
the lower line redshifts more than the upper line. The lower line is
now assigned to the quantum dot and the upper line to the cavity
mode. Because of the SCR, the eigenstates of the system are not the
exciton and photon anymore, but two exciton-photon mixed states or
dressed states, whose mixing depends on temperature. Let us
underline that these measurements are performed at low excitation
power, far below the QD saturation so that the mean photon number
within the cavity mode remains below unity. We observe the coupling
between an exciton and a single photon and the observed splitting is
the Vacuum Rabi Splitting.

\begin{figure}[h]
\epsfig{file=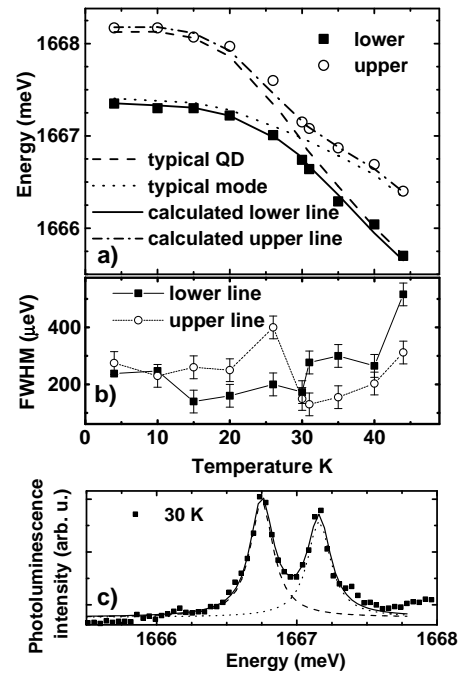,width=8 cm} \caption{\small {(a) Symbols
: emission energy of the upper and lower lines as a function of
temperature. Dashed (resp. dotted) line: typical spectral shift of
a quantum dot (resp. cavity mode). Continuous and dashed dot line
: calculated energy of the two coupled states. (b) Emission
linewidth of the lower (squares) and upper (circles) lines as a
function of temperature. (c) Squares : Emission spectrum at
   resonance showing the Rabi doublet. Dashed and dotted lines: lorentzien fit of each peak.
Continuous line: sum of the two lorentzien lines. } }
\end{figure}

 Figure 4a summarizes the emission energy of the upper and lower lines as a
function of temperature. We also reported the typical spectral shift
of a quantum dot (dashed line) and a cavity mode (dotted line). The
minimum value of the measured splitting between the two peaks
amounts to 400 $\mu eV $. Considering a coupling constant g = 200
$\mu eV $, we can calculate the energy of the two coupled states
using \cite {Fabre}:
\begin {eqnarray}
E_{\pm}(T) = \frac{E_{C}(T)+E_{QD}(T)}{2}\\
\pm \frac{1}{2}\sqrt{(E_{C}(T)-E_{QD}(T))^{2}+4g^{2}} \nonumber
\end {eqnarray}
where $E_{C}(T)$ and $E_{QD}(T)$ are the energy of the uncoupled
cavity mode and QD exciton as shown from fig. 4a. We obtain a very
good agreement with the experimental temperature dependance of the
lower and upper lines.

Figure 4c shows the measured doublet at resonance. Notice that the
two components have the same intensity. On the opposite, out of
resonance the signal is different for the quantum dot and for the
whispering gallery mode. This is due to their different emission
pattern within the microscope objective. Since at resonance, both
states are exactly half-exciton half-photon, they present the same
radiation pattern. As a result, the equal intensities we observe
within the doublet at resonance is another signature of the mixed
nature of the eigenstates. At resonance, the two lines of the
doublet present the same linewidth of 200 $\mu eV $.
 Theoretically, as both QD and cavity mode lines are homogeneously broadened, the linewidth of the mixed state is expected
  to be the sum of the exciton and photon linewidth
  weighted by the exciton and photon component of the mixed state. As temperature is increased, the photon like state
   (lower line below 30K, upper line                             above) linewidth remains unchanged and the QD-like state (upper line below 30K, lower line above) undergoes
   a thermal broadening \cite{Peterphonon}. At resonance, the
   linewidth of the two eigenstates
   is simply given by: $\frac{\gamma _{QD}+\gamma _{CM}}{2}$
where $\gamma _{QD}$ and $ \gamma _{CM}$ are the QD and cavity mode
linewith at the resonance temperature. As shown in fig. 4b, around
25K, the upper line (exciton like state) is thermally broadened and
then gets narrower around 30K because of the strong exciton-photon
mixing. This analysis in terms of mixed states explains the spectral
narrowing of the upper line observed around the resonance
temperature.

We now discuss the measured value of the Vacuum Rabi splitting. The coupling constant of the exciton-photon interaction
is given by \cite {Andreani} :
\begin {eqnarray}
g=\left(\frac{1}{4\pi \varepsilon _0 \varepsilon _r}    \frac{\pi e^2 f}{mV}\right) ^\frac{1}{2}
\end {eqnarray}

 $ f$ is the exciton oscillator strength, $V $ the effective modal volume, $m$ the free electron mass, $e $  the electron
  charge
 and $\varepsilon  _0 \varepsilon _r $ the dielectric constant. With a 2 $ \mu m $ diameter microdisk,
 $V \sim 6 (\frac{\lambda }{n})^3$, where $n$ is the effective refraction index of the cavity and $\lambda $ the
 emission wavelength. With the experimental Rabi splitting, we deduce
 an exciton oscillator strength of $f = 100$, a value comparable to values experimentally measured \cite {Stievater, Hourssubmit}. This
 value$ f = 100 $ is actually a minimum value since there may be a spatial mismatch between the QD and the optical mode
  antinode which would result in a smaller coupling constant. Then, to account for the measured splitting,
  larger oscillator strength would be required. As Andreani and coworkers have calculated \cite {Andreani} and as we have
  checked experimentally \cite {Hourssubmit},  the oscillator strength of monolayer fluctuation QDs strongly depends on their lateral size. From ref.\cite{Andreani},
  an oscillator strength of 100 corresponds to a lateral diameter of either 6 nm or 22 nm. According to the STM image of the
  QW interface, the QD under study most probably
  presents a diameter of at least 22 nm. Notice that  peak C and D on figure 2
    also present an anticrossing with a similar Rabi splitting, as temperature
is changed.   So in the  same microdisk, some QDs are in the SCR
with
  an optical mode whereas others are not: this can be due to variations of the oscillator strength from dot to dot
   and also to their position relatively to an antinode of the electromagnetic field.

To conclude, we have demonstrated the Strong Coupling Regime between
a large oscillator strength GaAs QDs and a high finesse microdisk
mode. This cavity geometry presents several promising advantages: in
addition to an easy fabrication process, extremely high quality
factors can be achieved associated to small effective volumes.
Recently we reported on the fabrication of similar microdisks
supported by an AlOx pedestal \cite {aplalox}. This pedestal makes
them very robust, avoid thermal heating and allows the fabrication
of all desired sizes in the same process. It is going to be
straightforward to insert GaAs QD in these microresonators and
achieve the SCR with even smaller effective volumes. Finally, if
several QDs are strongly coupled to the same optical mode, the QDs
can easily be selectively excited in this microdisk geometry, as
required for the manipulation of two qubits interacting via the
electromagnetic field \cite {Imamoglu}.
 The observation of the SCR between a single QD and a microcavity
  opens the way toward further fundamental
investigations of cavity quantum electrodynamics in a solid state system.

\begin{acknowledgments}
This work was partly supported by the "R\'egion Ile de France" and
the "Conseil G\'en\'eral de l'Essonne".
\end{acknowledgments}

\end{document}